\documentclass[prb,preprint]{revtex4-1} 
\usepackage{amsmath}  
\usepackage{amsfonts} 
\usepackage{graphicx} 
\usepackage{braket} 
\usepackage{xcolor} 
\usepackage[T1]{fontenc} \usepackage{ae} \usepackage{aecompl}
\usepackage[utf8]{inputenc}
\usepackage[hyperfootnotes=false]{hyperref}
\usepackage{rotating}
\usepackage{subfig}

\usepackage[]{footmisc}

\newcommand{\outcomment}[1]{}
\usepackage{caption}
\begin{document}


\title{Quantum Composer: A programmable quantum visualization and simulation tool for education and research}

\author{Shaeema Zaman Ahmed}
  \thanks{These three authors contributed equally.}
  \author{Jesper Hasseriis  Mohr Jensen}
  \thanks{These three authors contributed equally.}
  \author{Carrie Ann Weidner}
  \thanks{These three authors contributed equally.}
  \author{Jens Jakob Sørensen}
  \author{Marcel Mudrich}
  \author{Jacob Friis Sherson}
  \affiliation{Department of Physics and Astronomy, Aarhus University, Ny Munkegade 120,
8000 Aarhus C, Denmark} 
 
\date{\today}
\begin{abstract} 

Making quantum mechanical equations and concepts come to life through interactive simulation and visualization are commonplace for augmenting learning and teaching. 
However, graphical visualizations nearly always exhibit a set of hard-coded functionalities while corresponding text-based codes offer a higher degree of flexibility at the expense of steep learning curves or time investments.
We introduce Quantum Composer, which allows the user to build, expand, or explore quantum mechanical simulations by interacting with graphically connectable nodes, each corresponding to a physical concept, mathematical operation, visualization, etc.
Abstracting away numerical and programming details while at the same time retaining accessibility, emphasis on understanding, and rapid feedback mechanisms, we illustrate through a series of examples its open-ended applicability in both introductory and advanced quantum mechanics courses, student projects, and for visual exploration within research environments.

\end{abstract}

\maketitle 
\section{Introduction} 

Visualization and simulation tools have been an important component in teaching and learning quantum mechanics.~\cite{QMS, Thaller, Physlet_2003, Zollman, Escalada, Antje_2015, Antje_2019, PhET_QM, QUILT, QuVis} These tools are widely used to augment student understanding of basic quantum mechanical concepts such as superposition, \cite{PERC_2020} tunneling, \cite{Wieman} time evolution of wave functions, \cite{Kohnle_2019,Singh, Singh_2011} and expectation values. \cite{Singh_2017} Ranging in utility from self-study to classroom supplements alongside lectures, textbooks, or assignments, these tools enhance student understanding in a multitude of ways. For instance, in the PhET, \cite{PhET_QM} QuILTs, \cite{QUILT} and QuVis \cite{QuVis} projects, the simulation platforms and learning tutorials have been shown to further student understanding and assist students with tackling difficulties and building mental models of quantum systems. \cite{PERC_2014, Antje, Kohnle_2019, Wieman, Singh_2019}

Most of the current simulation platforms consist of a suite of pre-made modular simulations with an interactive experience.
Taking QuVis as an example, each simulation focuses on a single concept and includes a “Step-by-Step Exploration” tab that provides in-depth details.~\cite{QuVis}
Users typically interact with the tool by clicking icons and manipulating parameters through sliders, radio buttons, play controls, and check boxes. This enables and encourages exploration and investigation, \cite{QuVis} albeit such exploration is restricted to the static confines of the particular application.
Since continued development of software is often constrained or infeasible for a multitude of reasons (e.g. financial limitations, lack of personnel or time, or a low return on investment), expanding outside of any preconceived boundaries would usually require individuals to enter the arena of text-based programming environments. However, prerequisite knowledge of basic programming concepts, syntax, and language-dependent details presents a major roadblock for novices, in most cases placing such extension outside feasible reach.\cite{Alexander} 

In this paper, we present Quantum Composer, an interactive and flexible node-based tool for teaching and learning quantum mechanics. With certain similarities to other visual-based programming tools like LabVIEW~\cite{LabVIEW} and Scratch,~\cite{Scratch} the design distills physical and mathematical objects and concepts into intuitive, individual nodes. The user-defined inter-connectivity of these nodes then defines the behavior of the simulation. 
The large vocabulary of nodes allows numerous opportunities for investigation within a unified framework, with accessibility underscored by emphasis on drag-and-drop models and vivid visualizations. This allows students, instructors, and researchers to explore, expand, or even build from scratch \textit{flow-based} simulations of arbitrary complexities without prior programming knowledge.
We present how we have applied Quantum Composer from the high school level up to the graduate level as an educational tool, and we further highlight its potential as a research-assisting tool. That is, the simulations and visualizations allowed by Composer can allow users to explore research-relevant scenarios without the explicit need for deep programming expertise. This can serve both to lower the barrier of entry for students into research environments and provide experts with a quick and easy way to test ideas before applying them in more complex theoretical or experimental scenarios.

The paper is organized as follows: In Section 2, we give a brief overview of Quantum Composer, where we primarily describe the interface, features, and technicalities of the tool. In Section 3, we discuss how the tool can be used in an educational setting, and we present some example exercises in Section 4. 
Section 5 describes how we have used Quantum Composer as a research tool.
Finally, Section 6 concludes and provides an outlook. 

\section{Overview of Quantum Composer}

Quantum Composer (abbreviated as Composer) is an interactive tool for simulating and visualizing quantum mechanical concepts and systems. It consists of a graphical user interface where a user connects nodes to create a \textit{flow-based} simulation. The backbone of the tool is the QEngine, our numerical C++ library for quantum simulation and quantum optimal control. \cite{QEngine,Twoparticle,QM2} Composer allows users to explore a subset of the QEngine's capabilities: one-dimensional single-particle (Schr\"{o}dinger equation) and mean-field Bose-Einstein Condensate (Gross-Pitaevskii equation) physics,\cite{BEC-GPE} both in time-dependent and time-independent settings. Fundamental quantum mechanical calculations such as operator application, expectation values, operator variance, density integrals, state overlaps, and so forth are readily available. As a more advanced feature, the tool allows users to solve quantum optimal control problems using state-of-the-art gradient-based optimization algorithms.\cite{QEngine}

\subsection{Installation and documentation}

Quantum Composer is currently available as a free download for Windows, macOS, and Linux at \url{www.quatomic.com/composer/}. Basic functionality is explained in the \textit{Tutorial} pages (\url{www.quatomic.com/composer/tutorial/}). The extensive \textit{Reference} pages (\url{www.quatomic.com/composer/reference/}) provide detailed explanations and basic usage guides for all available nodes, as well as a \textit{Frequently Asked Questions} section that can help users understand and debug errors. For clarity of presentation, the overview provided in this section describes a relatively small subset of nodes.

\subsection{Interface and Features}
\begin{figure}[t]
    \includegraphics[width=\linewidth]{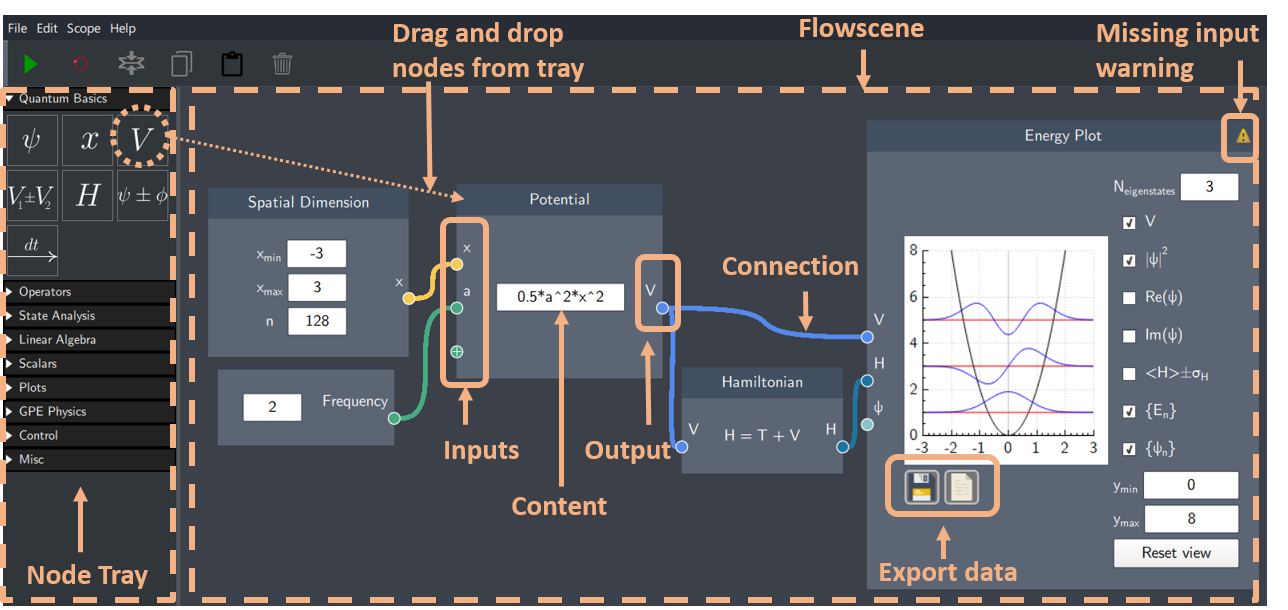}  
    \caption{Quantum Composer example flowscene showing how to plot the energy spectrum of a harmonic oscillator potential.
    Individual nodes can be dragged and dropped from the node tray (left dashed box) into the flowscene (right dashed box), which constitutes the simulation environment.
    The highlighted tray icon in the unfolded \textit{Quantum Basics} category shows its correspondence to the node in the scene.
    A simulation is assembled by connecting the nodes' inputs and outputs by dragging and dropping. A non-critical warning is displayed because one of the optional inputs to \textit{Energy Plot} is missing.}
    \label{Interface}
\end{figure}

In designing the interface of Quantum Composer, a `drag and drop' paradigm has been the focal point to create an accessible, easy to use, and flexible simulation environment. This kind of interface has shown to be intuitive and useful for educational simulations.\cite{Adams} A simulation in Quantum Composer is called a \textit{flowscene} and consists of a collection of interactive \textit{nodes} and \textit{connections} between these. 
Flowscenes can be saved to or loaded from a text-based \textit{.flow} file format, allowing simulations to be readily shared and distributed.   
An example of a flowscene is shown in Fig.~\ref{Interface}.
Usually, specialized \textit{Plot} nodes are used to visualize the result(s) of the simulation.
The content of these nodes can be saved as an image file, and their data can be exported as a \textit{.csv} file.

\begin{figure}[t]
    \includegraphics[width=\linewidth]{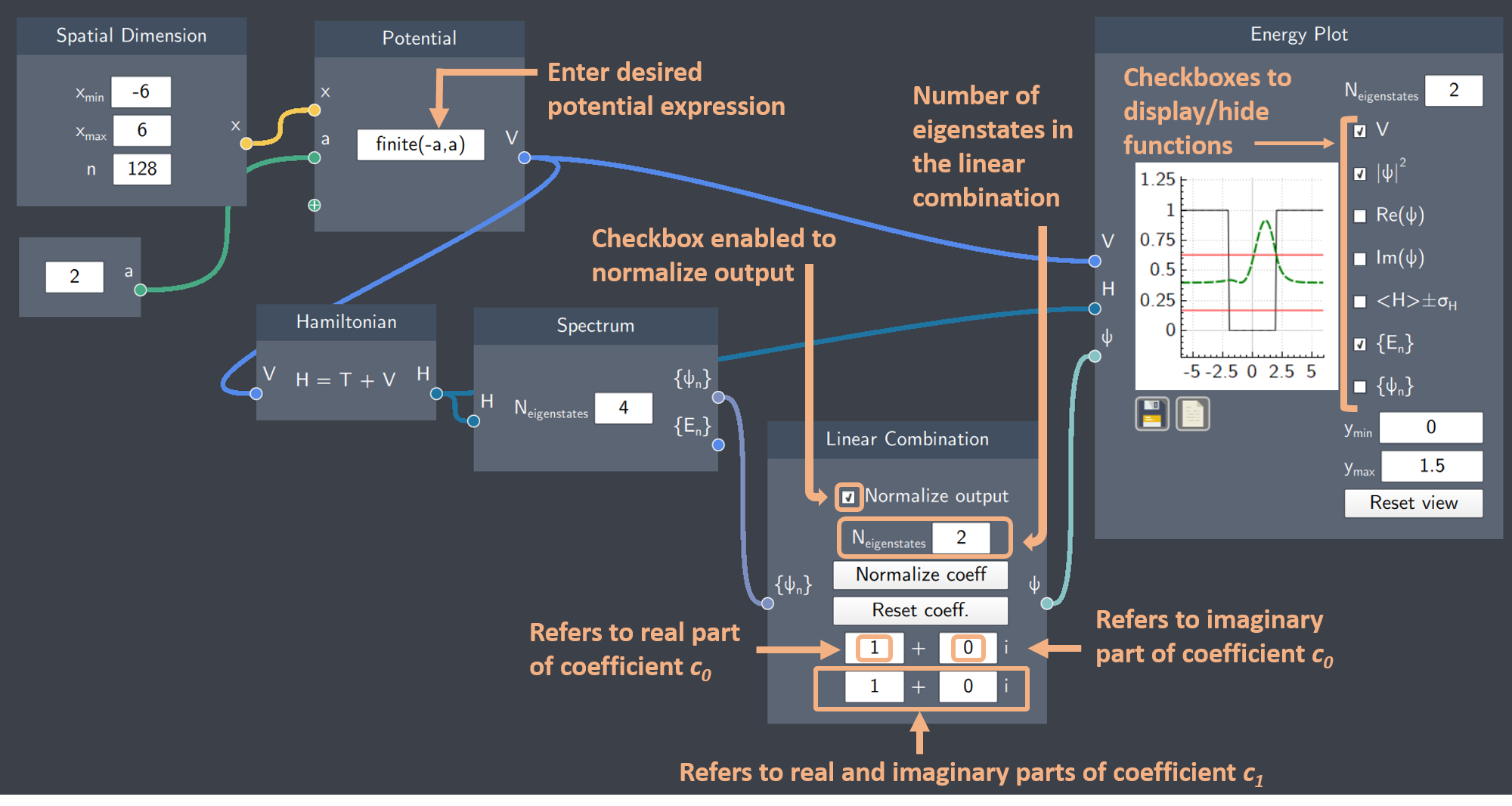}
    \caption{Screenshot of a flowscene explaining a few nodes. 
     In the \textit{Potential} node, the user can enter any desired potential expression, in this case a finite well of unit depth and width defined by the input scalar \textit{a}. A list of commonly used expressions are mentioned in the \textit{Reference} pages. The \textit{Linear Combination} node is used to create a linear combination of input states,
    here a superposition of the ground state and first excited state of a finite well $\psi_\mathrm{lin}(x) = c_{0}\psi_0(x) + c_{1}\psi_1(x)$, by entering values for the expansion coefficients in the indicated text-fields. Here, $c_0 = c_1 = 1$. The \textit{Energy Plot} shows the energies of the corresponding eigenstates of the Hamiltonian as well as the probability density of $\psi_\mathrm{lin}$ (offset on the y-axis by its associated energy). Several plotting options are available through the check boxes. This gives the user the ability to choose the amount and type of information they want to see simultaneously.}
    \label{fig:Node_Plot}
\end{figure}

All nodes consist of a title, contextual content, and a number of inputs and outputs.
A particular node is created by the user by dragging and dropping the corresponding icon from the node tray
(divided into categories of conceptual or functional similarity, e.g. \textit{Quantum Basics}, \textit{Operators}, and \textit{Plots})
onto the flowscene (the simulation environment workspace). In the flowscene, the output of one node can be connected to the input of another node by dragging and dropping either connection anchor onto the other, provided their types are the same (denoted by the color). Another example where the potential has been changed from the (default) harmonic oscillator to a finite well is shown in Fig.~\ref{fig:Node_Plot} where a flowscene and three of its nodes are elaborated in detail describing the inputs, output, and contextual content.

The functionality of a node is defined by how its contextual content operates on the provided input connections: once a connection is established between a pair of nodes,
data \textit{flows} from the output of the first into the input of the second (hence the term \textit{flow-based}) where it is read-accessible for further computation.
\footnote{Nodes are similar to functions in a functional programming paradigm} 
The computation only takes place if the necessary mandatory connections have been made, otherwise the node will indicate an error state in its top right corner.
If an optional connection is missing, a warning state is indicated instead, as explained in the \textit{Reference} pages.
This computation may yield an output that can be used as an input to a third and fourth node, and so on. The nature of the node connections provides explicit visualization of the dependency relations between different parts of the simulation, a feature not necessarily reproduced immediately (or at all) in text-based programming environments.

\begin{figure}[t]
    \centering
    \includegraphics [width=\linewidth]{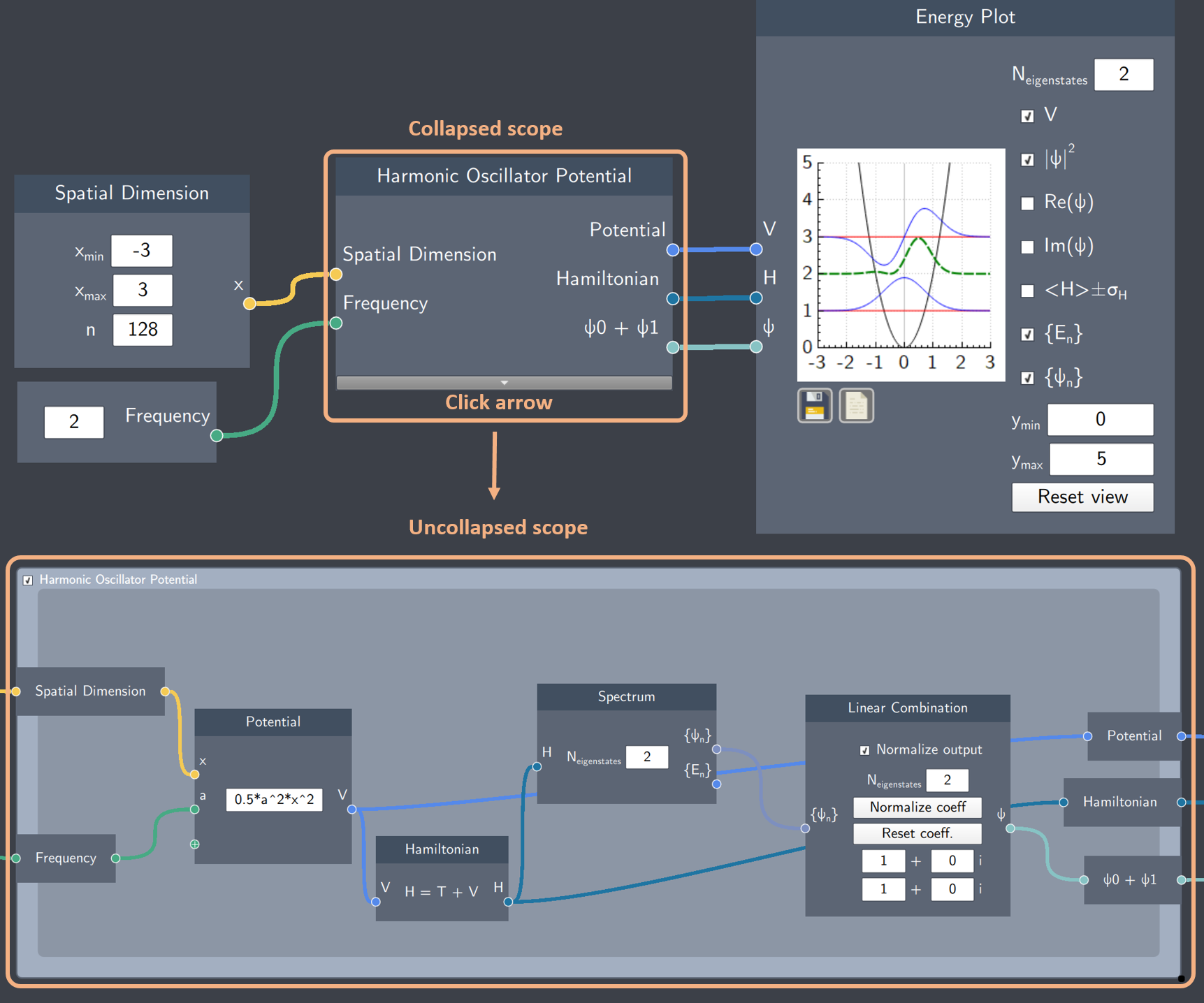}
    \caption{Nodes can be hidden by placing them in a \textit{Scope}, defining inputs and outputs, and collapsing the \textit{Scope}. In this example, an extension of Fig.~\ref{Interface} is shown and a \textit{Scope} node is used to hide certain nodes. In the collapsed setting, users only see the ``Frequency'' parameter and its effect on the resulting energy spectrum on the \textit{Energy Plot}.
    }
        \label{fig:Scope}
\end{figure}

If a flowscene has many nodes, the \textit{Scope} node can be used to compartmentalize any number of them, allowing customizable layers of abstraction as shown in Fig.~\ref{fig:Scope}.  A \textit{Scope} is created either from the node tray or by clicking and dragging the mouse cursor while pressing the
\verb|ctrl| or \verb|cmd| key on Windows/Linux and macOS systems, respectively. Node groups can then be hidden by dragging and dropping them into the \textit{Scope} node, setting up the internal connection logic, dragging connections across the \textit{Scope} boundaries to create inputs and outputs and finally \textit{collapsing} the \textit{Scope}. In Quantum Composer, any label (input, output, or \textit{Scope} title) can be customized by double-clicking it and typing out a new label name. This is especially useful for \textit{Scope} outputs since it allows simple, implicit documentation about the internal computation. A collapsed \textit{Scope} functions exactly like any other node, allowing the user to build new, customized nodes of arbitrary complexity, behavior, and level of abstraction from the pre-defined nodes. This is usually used to control or limit the amount of visible information, as shown in Fig.~\ref{fig:Scope}, with the purpose of reducing the cognitive load on a user or if an instructor would like students to interact only with a limited number of nodes in an educational setting. This constraint can allow students to interact only with the important part of the simulation in order to foster the development of their intuition.~\cite{Adams_2009}

\begin{figure}[t]
    \centering
    \includegraphics[width=\linewidth]{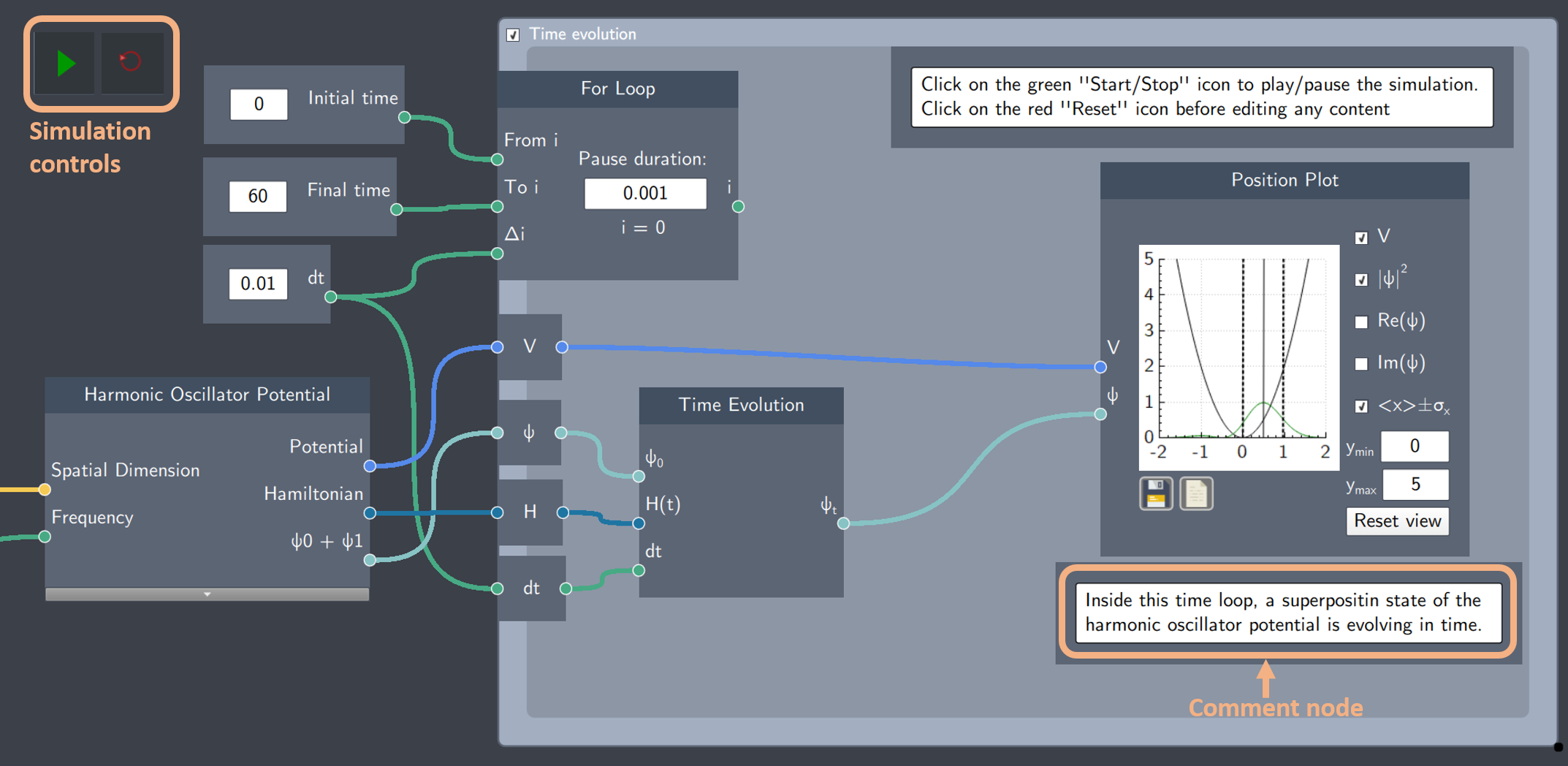} 
    \caption{Time evolution \textit{Scope} for real-time visualization of evolution of the superposition state from Fig.~\ref{fig:Scope} in the harmonic oscillator potential. The nodes in the \textit{Scope} (labelled \textit{Harmonic Oscillator Potential}) outside the \textit{For Loop} scope define the initial conditions of the time dynamics observed inside the scope.}
    \label{Time_evolution}
\end{figure}

To perform time evolution of a state, the \textit{Time Evolution} node must be placed inside a \textit{For Loop}, a special kind of \textit{Scope}. 
Fig.~\ref{Time_evolution} shows how to set up the time evolution of a superposition state in a harmonic oscillator potential by extending the flowscene shown in Fig.~\ref{fig:Scope}.
The \textit{Time Evolution} node must be provided a \textit{Hamiltonian} and a scalar time step as inputs along with an initial state, which could come from, e.g., a \textit{Linear Combination} or \textit{Analytic Wave Function} node. 
The user defines the time resolution, speed and duration of the dynamics. Additional nodes and plots can be inserted in the \textit{For Loop} scope to visualize different aspects of the dynamics like, for example, the density distribution of the wave function or various expectation values. After the simulation has been set up, the time evolution can be observed by clicking the green \textit{Play} icon, and a simulation can be modified after running by clicking the red \textit{Reset} icon. 
A time-dependent \textit{Potential} and \textit{Hamiltonian} can also be programmed by placing the nodes inside the \textit{For Loop} scope, and the \textit{For Loop} boundary node scalar output \textit{i} (corresponding in this case to the time variable) can be used as input to the potential.
Note that the \textit{For Loop} is generic and not limited just to time evolution, although this is the most frequent use case. Like a regular \textit{Scope}, a \textit{For Loop} can also be collapsed, and any outputs are evaluated only at the end of the loop. Quantum Composer also has a \textit{Comment} node where one can add text boxes anywhere in the flowscene in order to make a simulation with self-contained explanations. These are similar to inline comments in text-based programming.

\begin{figure}[t]
    \centering
    \includegraphics[width=\linewidth]{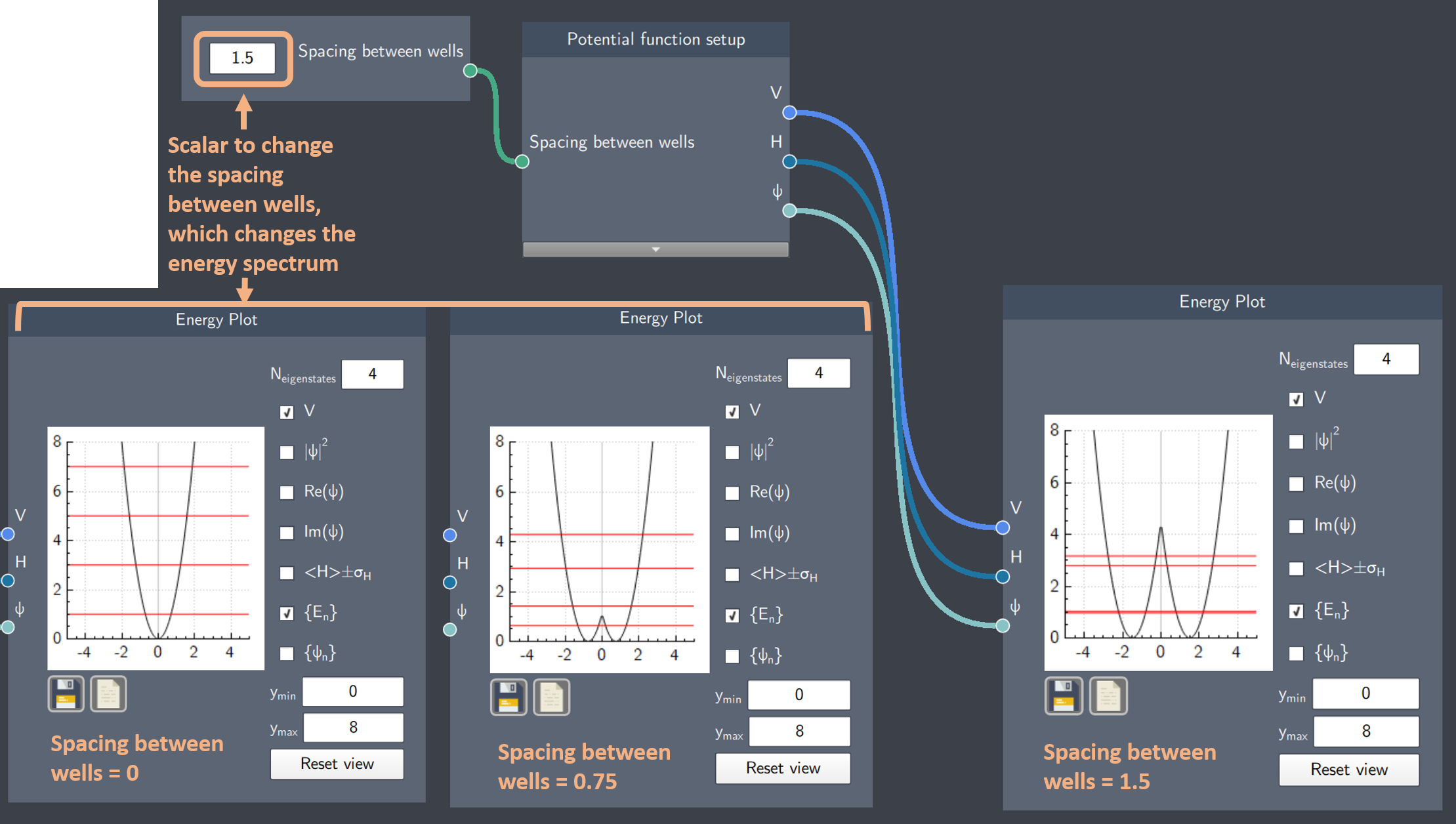}
    \caption{Changing from a single-well to a double-well potential by entering real scalar values in the \textit{Spacing between wells} parameter. Changing this parameter results in splitting of the wells. The energy levels of the system also split to create energy bands and bandgaps, which can be instantaneously seen in the \textit{Energy Plot}. The \textit{Scope} node abstracts away most of the implementation details, greatly reducing the user's cognitive load.}
    \label{fig:double well}
\end{figure}

In Quantum Composer, an instantaneous update of the flowscene is triggered when either the contextual content or input of a node is altered. The new changes cascade through all dependent connections. This enables swift, high-level behavior in a very accessible platform. As a practical example, the change of the energy spectrum from a single-well to a double-well system can be visualized rapidly by just changing a single parameter governing the well separation, as shown in Fig.~\ref{fig:double well}, while the exact implementation details remain hidden in a \textit{Scope}. As the separation barrier is raised the increasingly degenerate energy levels bunch in pairs corresponding to symmetric and anti-symmetric states. Studies show that this kind of rapid response helps users to explore, iterate, and understand the governing factors that affect their simulation and also help to explore cause-and-effect relationships.~\cite{Adams_2008}

\section{Quantum Composer as an educational tool}

For the past three years, Composer has been used to teach introductory quantum mechanics at multiple universities around Europe. At Aarhus University, we have used Composer in such introductory classes for both physics and nanoscience majors as well as graduate-level quantum mechanics courses for physics students. In the latter two cases, Composer visualizations were used in exams. Additionally, we have used Composer to introduce high school students to quantum mechanics in focused workshop settings. Below we describe the use of Composer both as a classroom and an outreach tool. Finally, we describe a specific example of a workshop where university students were taken through a rapid ``mini-research project'' guiding them through all of the steps of the research cycle in preparation for a larger research project.

\subsection{Composer in the classroom}

Quantum Composer can be used by instructors and students to teach and learn quantum mechanics. In this section, we present some of the ways the tool can be used in an educational setting:
\begin{itemize}
    \item \textbf{Visualization:} Composer can be used to visualize a quantum mechanical concept or system for students during a lecture. For example, students have been shown to face difficulties in sketching the shape of the wave function even if their mathematical form of the wave function is correct.\cite{Singh}  Students also struggle with correctly describing the time development of a non-stationary state.\cite{Singh,Singh_2011} Quantum Composer can be used to visualize these concepts and connect them with the mathematics students learn during a classroom lecture.
    \item \textbf{Exploration:} Composer can be used for exploring different kinds of quantum systems that are not necessarily covered in their curriculum. For example, students are often taught the infinite well, the harmonic oscillator, and the free particle. Analytic solutions of other, more complex systems (e.g. the quartic oscillator) become rapidly untenable, especially in introductory-level courses. Quantum Composer can be used to investigate any potential numerically which gives students a means to visualize how quantum systems behave without solving them analytically. The idea is that by exploring these systems, students can build up their understanding of the underlying physics. For example, students can enter any arbitrary potential and visualize the resulting energy spectrum. Open-ended student exploration can range from building a simulation from scratch to investigating a system already defined by the instructor (e.g. by loading a pre-made simulation).
    \item \textbf{Student-driven discussions:} Instructors can use the tool in a whole-class setting by showing a scenario in Quantum Composer and asking students to predict the effect of manipulating variables (cf. Fig.~\ref{fig:double well}). Students can discuss their predictions in groups, and these can then be explored in Composer. As the simulation can be edited in real-time, giving immediate feedback to the students, new suggestions and inquiries from the students can sustain an ongoing inquiry-based discussion. 
    \item \textbf{Assignments:} Instructors can use Quantum Composer to prepare stand-alone assignments or couple them with theoretical assignments. At Aarhus University, we have developed stand-alone, curriculum-based exercises with pre-made flowscenes. These exercises typically use a guided-inquiry approach where students explore dependencies between parameters and quantities. The exercises can be downloaded from \url{https://www.quatomic.com/composer/exercises/}, and some of them are discussed in detail in Sec.~\ref{sec:exercises}.
    \end{itemize}

\outcomment{
\begin{figure}[t]
    \centering
    \fbox{
    \includegraphics[width=\textwidth]{Figures/Assignment_snapshot.PNG}
    }
    \caption{Example assignment question about superposition and expectation values}
    \label{Assignment}
    
\end{figure}
}

\subsection{Composer for dissemination and outreach}

Due to its inherent flexibility, Composer can be used to illustrate and visualize complex quantum mechanical concepts. These features can be used to facilitate dissemination and outreach projects that allow expert researchers to communicate their ideas and methods to students at a more introductory level.

As an example, Composer has been used by some of the PhD students of an EU Horizon 2020 project: QuSCo.~\cite{QuSCo} The QuSCo network consists of early-stage researchers (and their associated PIs) working in different domains of quantum control and sensing ranging from superconducting qubits, NV centers, ultracold atoms, and theoretical quantum optimal control. In a workshop setting, we mentored the PhD students from the QuSCo network as they simulated a component of their research using Composer.

As a part of the workshop, the students created a video explaining their simulation in language understandable to students studying introductory quantum mechanics. For example, a student working with qubits simulated Rabi oscillations, and their simulation was then explained briefly in a one-minute video and uploaded to the QuSCo website (\url{http://qusco-itn.eu/esr/}). The library of simulations created after this workshop could then be used to expose students to cutting-edge quantum physics research. The workshop had the additional benefit of training the PhD students to communicate their research in an engaging way.

\subsection{Composer workshops for students}

Composer can be used to structure a self-contained, workshop-style curriculum that instructs on aspects of quantum mechanics relevant to the individual setting. 
As an example, we have used Composer to build an introductory workshop on quantum control and optimization. This workshop is given to all new students and interns joining our quantum research group, and the only prerequisite is a course in introductory quantum mechanics. In this workshop, students are given a brief ($\approx 30$ minute) onboarding presentation, then they spend the next $20-40$ hours working through the material.

While the example we provide here is specific to our group, this workshop can be modified to explore relevant aspects of research done in other groups. The overarching goal of the workshop is to have students quickly work through each of the phases of a research project, e.g. hypothesis formulation, simulation design, data generation and analysis, and communication and discussion of final results.

In our workshop, students initially work through an initial set of self-guided exercises (including the spectra and correlation amplitude exercises described in Section~\ref{sec:exercises}). These exercises allow students to reinforce their understanding of quantum mechanics, but in addition they are, throughout the course of the workshop, introduced to concepts in numerical quantum computing and quantum control. In each case, the concept introduction requires students to read some background material and then work through visualization exercises in Composer. In all cases, students are encouraged to ask questions, and in our experience, e-mail availability is sufficient to answer these questions.

We then provide students with an open-ended challenge in which they try to solve a question that has been relevant in modern research in ultracold atomic physics: the excitation of a Bose-Einstein condensate (BEC) from the ground state to the first excited state.\cite{Schmiedmayer, QM2} Students are presented with the option to control a non-interacting single particle or an interacting BEC (where interactions are governed by the nonlinear Gross-Pitaevskii equation).\cite{Pethick} They are also given the option to work with a harmonic or anharmonic potential. Students are given suggestions as to what they can explore, but the specifics of what they choose to do is left up to them. The final product is a $2-4$ page report at the end of their exploration in a \LaTeX~template we provide.\footnote{Students are given the common revtex template in two-column format. The hope is that students feel that their report resembles that of a real research paper.} For example, students can explore why some state-to-state transfers of a single particle are not possible for a harmonic potential but can be done in an anharmonic potential like the quartic oscillator.

After the final reports are submitted, we read them and provide concrete feedback. In this way, students are trained in the critical thinking, organization, and communication aspects necessary to undertake a subsequent larger research project. This exercise also allows us to identify areas where students face difficulties and thus we can focus our efforts on strengthening their competences as they work through their larger project.

\section{\label{sec:exercises} Example exercises}

We have developed a set of exercises using Quantum Composer targeted at students from the advanced high school level through to the graduate level. These can be found at \url{www.quatomic.com/composer/exercises/} alongside related \textit{.flow} files and descriptions. In this section, we first present a brief description of the onboarding process and then highlight some example exercises at different levels to show how they could be used in-class and outside of class.

\subsection{Onboarding}

As in the student workshop described in the previous section, exercises are usually not deployed without any introductory onboarding to Composer. The aim of the introduction is to expose students to the Composer interface and take them through some simple examples. We typically ask students to install Composer on their laptops before coming to the session.\footnote{We often mention to students that they may prefer to use a mouse with Composer, especially those using macOS.}

In-session, we project a laptop screen so all participants can follow along, and we encourage students to work along with us. Then, we provide a brief overview of the Composer window and explain where students can find the different nodes. However, this part of the session should be brief, as we find it is best when students remain engaged. Therefore, after this broad overview, we begin by building a simple flowscene from scratch to gain familiarity with basic Composer usage through examples. Typically, we simply begin by setting up the flowscene that students will be working within the exercises that follow. The specific flowscenes that we set up therefore vary from session to session. However, we try to (at least) show students how to visualize potentials, eigenstates, and eigenenergies. If necessary, we also show students how to set up time evolution in Composer. However, we have found that sometimes an additional onboarding session is needed to introduce students to time evolution. This is especially true if time evolution is not used in the first exercise they are given but is present in future exercises. After onboarding, students then begin to work through the exercises while a Composer-competent instructor is present. This allows students to ask questions and receive answers in real-time, and we have experienced that this is helpful in helping students over the initial learning curve required to use Composer. A tutorial video illustrating these concepts can be found at~\url{https://www.quatomic.com/composer/tutorial/}.

It is important to note that, while we typically aim to equip students with the tools necessary to build their own flowscenes, we always provide students with ready-built \textit{.flow} files that they can use in the exercises. This allows interested students to use the ready-built flowscenes contained in the files as a reference if they get stuck building their own flowscenes, and others can simply use them to do the exercise without needing to build the flowscenes themselves. We recommend this when using Composer in practice, as it gives students the most flexibility in what they choose to explore, but they always have the most important visualizations readily available.

\begin{figure}[t]
    \centering
    \fbox
    {
    \includegraphics[width=6.5in]{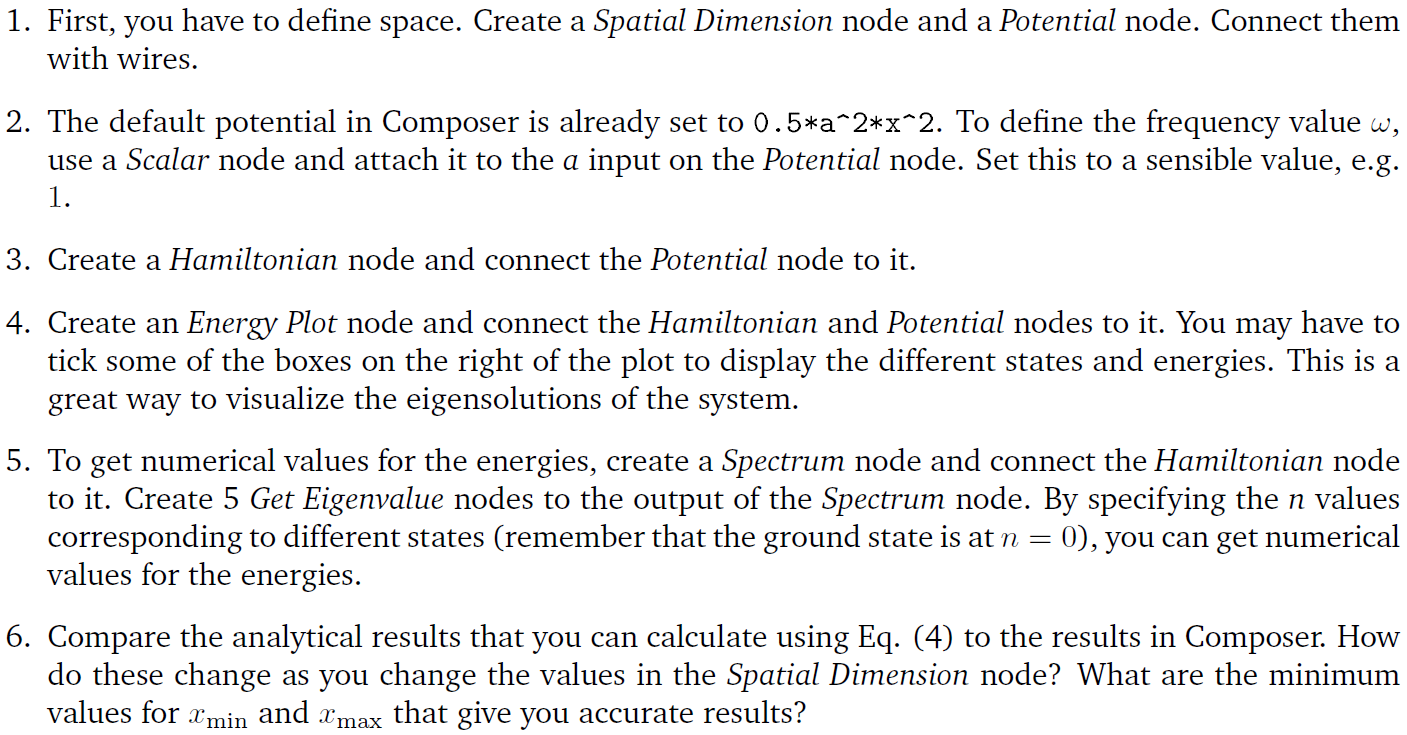}
    }
    \caption{A snippet from an exercise where students are asked to construct and explore the quantum harmonic oscillator potential. Step-by-step instructions are given to the student that they can use to build the flowscene themselves, and the instructions are worded in such a way that the student understands why the program is designed as it is.}
    \label{fig:text_onboarding}
\end{figure}

We are of the opinion that formal onboarding is always best, as it allows us to be physically present when students work through any initial understandings. In the absence of formal onboarding, we have found that students can still learn and use Composer without an instructor physically present. In these cases, however, detailed explanations that take students through the setup and execution of the flowscene are necessary. An example of this is shown in Fig.~\ref{fig:text_onboarding}. Note that step six in Fig.~\ref{fig:text_onboarding} asks students to explore the pitfalls that one may encounter when using  Composer (e.g. unexpected results due to finite numerical resolution, the non-periodic boundary conditions when calculating the spectrum, and so on). We recommend that students be made aware of the numerical limitations of Composer upon onboarding,\footnote{This is described in more detail at \url{https://www.quatomic.com/composer/reference/faq/}.} as well as the fact that, in this program, the units have been made dimensionless so that $\hbar = m =1$ for some mass $m$.\footnote{For example, to translate from Composer units to units used, e.g. in our quantum gas microscopy experiment, we take $m = 1$ to correspond to the mass of one Rb-87 atom and and energyscale $E$ to correspond to one photon recoil for a lattice with wavelength $532$~nm. For more information, see \url{https://www.quatomic.com/composer/reference/quantum-composer-basics/units/}.} We have found that such text onboarding can be time-consuming to construct, but with it students have been able to successfully utilize Composer, given that an instructor is virtually available to answer questions via chat or e-mail. Students found it helpful in some cases to attach screenshots, especially when debugging programming errors in Composer.

\subsection{The Exercises}

The exercises we have created span basic topics like superposition through to advanced topics like quantum optimal control of Bose-Einstein condensates, and we highlight some of the concepts illustrated by the exercises in this subsection. Each exercise mentioned here can be found in the online repository mentioned in the introduction to this section.

\begin{figure}[t]
    \centering
    \includegraphics[width=6.5in]{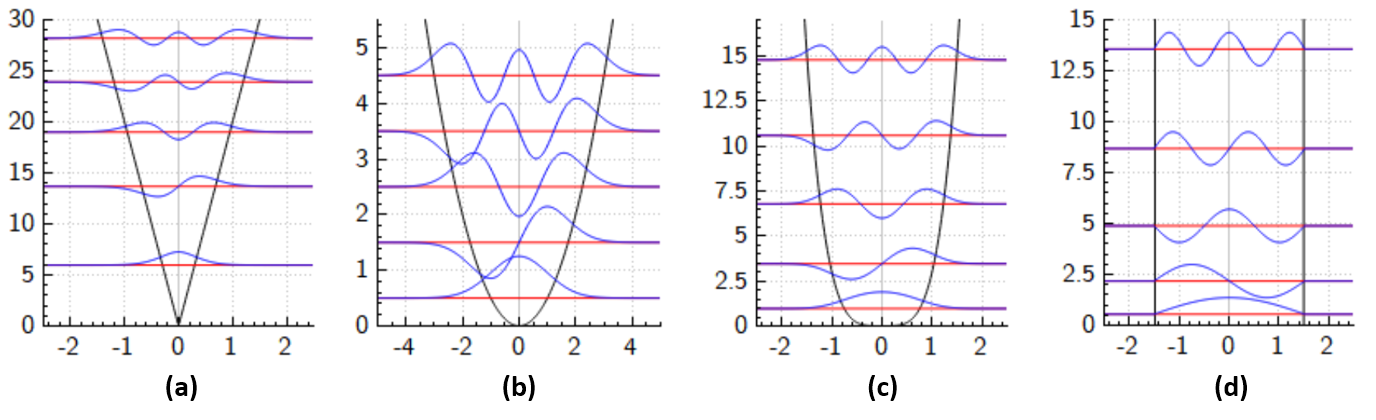}
    \caption{Spectra for the (a) absolute-value, (b) harmonic, (c) quartic, and (d) infinite well potentials, showing how the energy spacing $E_n$ goes from sub-linear, to linear, to sub-quadratic, to quadratic in $n$. The x-axis is in scaled position units, and the y-axis is in scaled energy units.}
    \label{fig:spectra}
\end{figure}

Basic exercises can be used to illustrate different aspects of visual quantum mechanics, like how harmonic oscillator energy eigenstates contain progressively more nodes as the principal quantum number increases (as can be seen in Fig.~\ref{Interface}). Students can also visualize both how energy eigenstates and levels change as potential parameters or the potential expression itself changes, as shown in Fig~\ref{fig:spectra}. Such an exercise in particular (called ``Spectra'' on the Exercises page of the website) has been adapted for students of all levels, and we find it is a useful way to help students build intuition of what happens as one moves from, e.g., a harmonic potential to a square well potential. That is, by going through the exercise, we aim to show how the spacing of the energy levels $E_n$ in the different systems moves from being sub-linear in $n$ to quadratic in $n$ as one moves from an absolute-value potential to an infinite well potential.
The idea is that by directly visualizing these changes as the expression in the \textit{Potential} node changes, students can identify how changes in the potential affect the resulting energy spectrum (i.e. that the slope of the potential at the edges largely governs how the eigenvalues scale). The goal of this exercise is to help students build intuition about energy scaling in systems, including those not typically covered in the quantum curriculum. Such exploration could allow them to glean a better understanding of why the systems they study in detail behave the way they do.

\begin{figure}[t]
    \centering
    \includegraphics[width=5.2in]{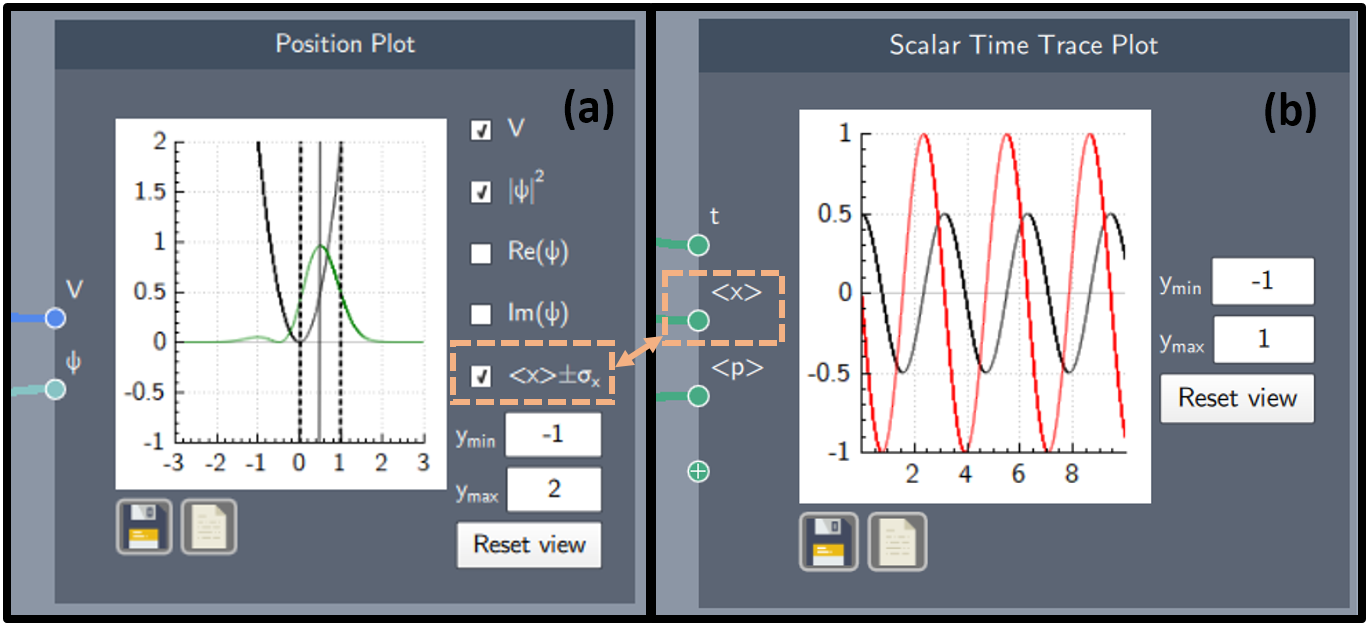}
    \caption{Given an initial state that is an equal superposition between the ground and first excited states of the harmonic oscillator potential (cf. Fig.~\ref{fig:Node_Plot}), the plots show (a) the potential (black parabola), the state density (green), $\braket{\hat{x}}$ (black vertical line), and $\braket{\hat{x}} \pm \sigma_x$ (black dashed lines) and (b) how $\braket{\hat{x}}$ (red) and $\braket{\hat{p}}$ (black) evolve in time (as indicated on the x-axis). The orange dashed boxes and arrows indicate two different ways to visualize $\braket{\hat{x}}$ in Composer.}
    \label{fig:time_ev_exp}
\end{figure}

More advanced exercises can illustrate concepts like the time-evolution of expectation values, as shown in Fig.~\ref{fig:time_ev_exp}, which is useful in an introductory quantum mechanics course. In particular, the expectation value of position can be shown in a \textit{Position Plot} and operators like position and momentum can be visualized by plotting these as time evolves in a \textit{Scalar Time Trace} plot. This allows students to visualize how states and their expectation values evolve in time.

\begin{figure}[h!]
    \centering
    \includegraphics[width=5.1in]{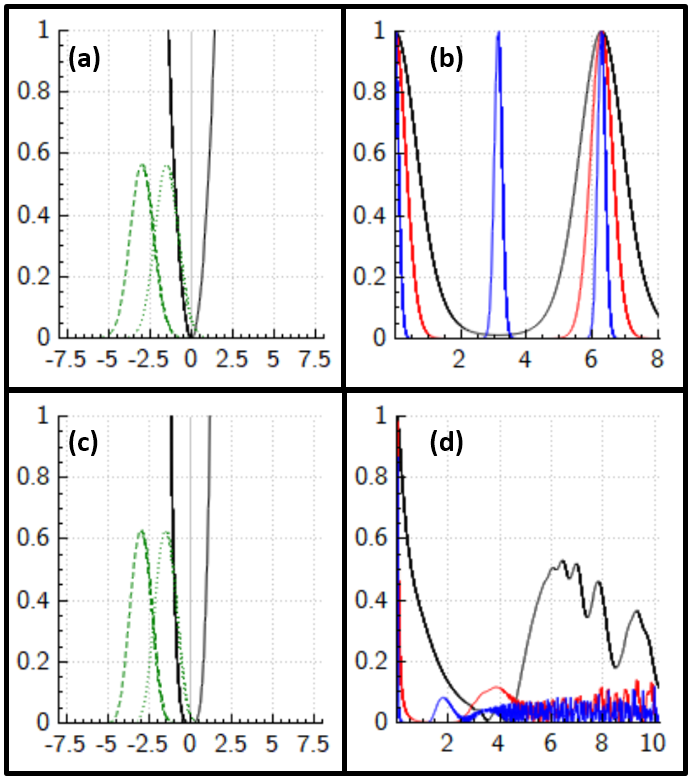}
    \caption{(a) Harmonic potential (black) and two of the displaced ground states ($x_0 = 1.5, 3$ for the dotted and dashed curves, respectively) used as initial states to plot (b) $|C(t)|^2$ in Eq.~\ref{eq:correlation} for ($\omega$, $x_0$, color) of (1,1.5, black), (1, 3, red), and (2, 3, blue). The two states shown in (a) correspond to the black and red traces in (b), respectively. 
    (c) Same as (a), but for the quartic (anharmonic) potential. (d) Same as (b), but for (1, 1.5, black), (1, 3, red), and (0.5, 3, blue).
    }
    \label{fig:correlation}
\end{figure}

In graduate-level settings, Composer can be used to visualize aspects like the norm square of the correlation amplitude
\begin{equation}
    \label{eq:correlation}
    |C(t)|^2 = |\bra{\psi(t = 0}\hat{U}(T,0)\ket{\psi(t = 0)}|^2
\end{equation}
between an initial state $\ket{\psi(t=0)}$ and the time-evolved state $\ket{\psi(t = T)} = \hat{U}(T,0)\ket{\psi(0)}$.\cite{Sakurai} We illustrate this in Composer by displacing the ground state of a harmonic oscillator, with $V(x) = \omega^2x^2/2$, from the potential center, as shown in Fig.~\ref{fig:correlation}(a). Composer then calculates $|C(t)|^2$ as a function of time. Students explore how the correlation amplitude decays and revives over time, and they are asked to quantify how this changes as a function of the state displacement (the variable $x_0$ in Fig.~\ref{fig:correlation}) and the frequency $\omega$ of the harmonic potential. Example traces of $|C(t)|
^2$ for three different parameter choices are shown in Fig.~\ref{fig:correlation}(b). 
Finally, students explore the same situation for the quartic (anharmonic) oscillator with $V(x) = \omega^2x^4/2$ in Fig.~\ref{fig:correlation}(c), where a breakdown of this revival behaviour is shown in Fig.~\ref{fig:correlation}(d).
\footnote{The rapid oscillations seen in Fig.~\ref{fig:correlation}(d) are physical phenomena due to wave function interference arising from multiple reflections at the edges of the anharmonic potential.}
This example serves to show how systems that initially appear to be very similar, as indicated by the visual similarities in the initial states and potentials in Fig.~\ref{fig:correlation}(a) and (c), indeed behave very differently due to differences in their spectrum (cf. the spectrum plots shown in Fig.~\ref{fig:spectra}(b) and (c) for the harmonic and anharmonic cases, respectively). 

These exercises represent a small portion of the possible explorations accessible in Composer. We present them as examples that span educational level, conceptual space, and complexity, and they serve simply to highlight the potential value Composer brings to quantum physics education.

\section{Quantum Composer as a research-assisting tool}

In this section, we present some of the scenarios where Composer has been or could be used as a research-assisting tool in the ultracold atoms field. That is, we present our (subjective) experiences with incorporating the tool's qualitative and quantitative capabilities into the different stages of a research project. In a certain sense, these have appreciable overlaps and parallels to \textit{repurposing} citizen science games,\cite{FoldIT,eyewire} which are inherently geared towards non-experts, into software useful for expert scientists.\cite{cooper2018repurposing} The primary notion, in this case, is that software artifacts (distinctly different from e.g. the produced data) and design principles may emerge in developing for the former audience that can prove valuable also for the latter. Among others, the core tenets (e.g. flexibility, ease of use, rapid iterability) of Quantum Composer reinforce said ideas and remain central also in this context as elaborated in the following examples.

\subsection{Theoretical research}
A substantial amount of our research is geared towards quantum optimal control, \cite{QEngine,QM2,Twoparticle,ExactDerivatives,GROUP1,GROUP2,AlphaZero} a mature, but evolving, field of research as it is one of the requisites for quantum technology experiments.\cite{Quantum_Tech} 
A common task in quantum optimal control is to steer an initial quantum state to a target state with desired accuracy in the most optimal way.\cite{QEngine} 
This is achieved by influencing the system dynamics through certain controls (e.g. the depth or center of an optical tweezer) that parameterize the Hamiltonian in a way specific to a given problem. 
Usually, \textit{optimal} controls are found by iterative, local optimization of many different (often order $10^2-10^4$) initial guesses, or seeds.
This paradigm, known as multistarted local optimization, typically involves large-scale parallel optimization processes on e.g. a computer cluster. 

Quantum Composer includes the capability to set up, solve, and analyze such state transfer problems.
Although not by itself suitable for advanced control techniques like multistarting, we have had very positive experiences with using the program both as a precursor for the simulation (problem definition), optimization, and analysis of the resulting optimal dynamics. 
As a concrete example, the control problem studied in Ref.~\onlinecite{Twoparticle} has a non-trivial dependence on some of the control parameters. We used Composer to e.g. 
(1) probe different regimes and combinations of these parameters to deepen our understanding of the problem, which is useful in part for designing seeding strategies,
(2) confirm that the geometry was correctly modelled by comparing the energy spectrum to a benchmark paper,
(3) find proper boundaries for the control values for keeping lattice unit cells unmixed (a central assumption), and 
(4) confirm unit conversions between simulation and SI units.
We used Composer for similar tasks in Ref.~\onlinecite{QM2}, which treats several single-particle and BEC control problems. There, we also estimated the timescale necessary for appreciable dynamics and used those results directly for related rough quantitative estimates included in the published paper.
Additionally, we have performed numerous smaller, internal numerical studies in a wide variety of different contexts, e.g. for studying how the transmission and reflection of Gaussian wavepackets impinging on different types of potential barriers changes with the initial spatial and momentum distributions.

In principle, the above investigations and visualizations could be conventionally conducted by writing corresponding C++ native source code (assuming prior installation of library dependencies), writing compilation scripts, building and linking the program, running the executable, exporting the results, importing the results in an appropriate tool, writing visualization code, and finally executing it. 
This multi-staged procedure, however, ranges from being quite tedious to unrealistic, depending on the individual user's programming experience and needs. Especially, even for the most basic applications, only a fraction of the time is spent on dealing with the actual physical concepts and phenomena under consideration.
Further, both programming and research are inherently fine-grained iterative processes,
each iteration changing often just a tiny bit of behavior, but nonetheless requiring invocation of the full pipeline. One of the defining core tenets of Composer, as highlighted throughout this paper, circumvents this chain by lowering the barrier for entry, containing the work space to a single application, allowing for rapid feedback on changes without re-building, and focusing on the visualization of physical concepts. 

Overall, we find that, rather than being a complete replacement for text-based programming, Quantum Composer is a welcomed and effective addition in the suite of available tools at our disposal.
 
 \subsection{Experimental research}
 
We have found that Composer is also a useful tool for exploring experimentally-relevant potentials and the dynamics of quantum states within these potentials, especially if one is just beginning to work in research. For example, many dipole potentials used for atom trapping are generated with Gaussian laser beams.\cite{Grimm} How these potentials and their energy eigenstates are affected e.g. by the beam waist or laser power is straightforward to model in Composer. It is also trivial to add the effects of gravitational sag when determining whether or not a trap can hold up against gravity. One can also explore and optimize atom dynamics in such traps. This is a necessary precursor to actual experimental implementation, as one can simulate a system and verify its utility before undertaking the often expensive and time-consuming process of laboratory setup and testing. Thus, easy-to-use tools like Composer allow the experimentalist to perform such tests without the overhead of most numerical programming methods.
 
In our experimental research group, we found Composer to be a practical and useful tool for determining whether or not single atoms trapped in an optical lattice potential could be prohibited from tunneling to certain lattice sites using an attractive dipole potential. This problem is typically solved in practice by using a repulsive potential to remove a well and replace it with a potential barrier. Intuitively, it seems strange to \textit{plug} a well to inhibit tunneling using attractive potentials. However, if one thinks of tunneling as the temporal evolution of a non-stationary state (e.g. a particle localized to a single well of a multi-well system), then it stands to reason that tunneling is inhibited if one shifts the energy states of a given well (or wells) out of resonance with the rest of the wells. The question remains, however, whether such potential shifting can be done without bringing higher-excited levels into resonance with the ground state of the un-shifted wells.
 
\begin{figure}[t]
    \centering
    \includegraphics[width=6in]{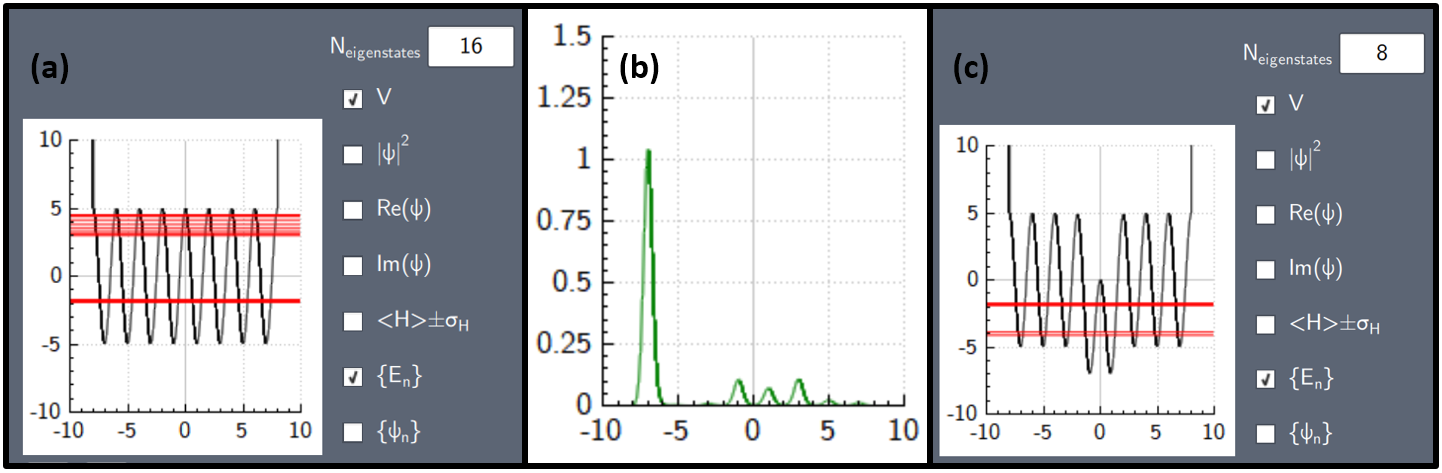}
    \caption{(a) \textit{Energy Plot} showing the $8$-well lattice potential (black) and the energy landscape for the first $16$ eigenstates (red lines), (b) \textit{Position Plot} showing the state (green) localized as much as possible to the first well, (c) Another \textit{Energy Plot} showing the lattice potential with the Gaussian plug, showing that the lowest band has been split, and two energies now sit below the original lowest band shown in (a).}
    \label{fig:lattice}
\end{figure}

To illustrate this in Composer, we set up an eight-well system as shown in Fig. \ref{fig:lattice}(a). This cosinusoidal lattice represents part of our (effectively) infinite optical lattice potential in which we trap single atoms. The lattice potential is given by 
\begin{equation}
\label{eq:lattice}
    V_\mathrm{latt}(x) = 
    \begin{cases}
    V_0\cos{(2kx)}, \quad \text{if} \quad -8  \leq x \leq 8, \\
    \infty, \; \text{otherwise}.
    \end{cases}
\end{equation}
Here, we include the hard wall boundaries to avoid any confusion due to edge effects, as Composer uses a combination of absorbing and periodic boundary conditions (cf. the \textit{Reference} pages on the Composer website). We define our spatial coordinate such that  $-10  \leq x \leq 10$. It is good practice to work with potentials that tend towards infinity at the boundaries to preserve numerical accuracy and physicality of simulations.

For the purposes of this simulation, the lattice depth is $V_0 = 5 E_\mathrm{rec}$, where the recoil energy is defined as $E_\mathrm{rec} = \hbar^2k^2/2m$ for a lattice with wavenumber $k = 2\pi/\lambda$, wavelength $\lambda = 1064$~nm, and the mass here refers to $m_\mathrm{Rb} = 1.44\times 10^{^-25}$~kg, the mass of the rubidium-87 atoms we trap in our experiment.\cite{QGM} Note that $E_\mathrm{rec} = 1$ in this simulation. The lattice depth is low enough that we do not expect tunneling to be fully suppressed due to the superfluid-to-Mott-insulating transition in one dimension (the other two dimensions are fully frozen out).\cite{1D_SF_MI} As we expect to populate only one atom per site when preparing our states, however, these simulations are single-particle and neglect any nonlinearity due to the fact that we load Bose-condensed atoms into the lattice. However, similar investigations  of BEC dynamics are possible in Composer.
 
The energy landscape of this potential is shown in Fig.~\ref{fig:lattice}(a), and we see that the first $16$ eigenstates make up two bands of energies trapped in the lattice. By creating an equal superposition of all states trapped in the first band ($8$ total eigenstates), we can create an approximation to a state localized in the first well of the lattice, as shown in Fig.~\ref{fig:lattice}(b). The ``leakage'' of this state into other wells is a consequence of the finite number of wells considered in this simulation.
To illustrate the effect of well \textit{plugging}, we superpose on this potential a Gaussian potential (representing a laser-created dipole potential) of the form
\begin{equation}
\label{eq:plug}
    V_\mathrm{plug}(x) = -A\exp{\bigg\{-\frac{(x-x_0)^2}{b^2}\bigg\}},
\end{equation}
i.e. the total potential is $V(x) = V_\mathrm{latt}(x) + V_\mathrm{plug}(x)$, where we set the amplitude of the plug to $A = 5$ and its width to be $b = 1$ (noting that in scaled units, each lattice site spans $\lambda/2 = 2$). If we center the plug potential around zero, $x_0 = 0$, we expect that wells $4$ and $5$ (counting wells $1,...,8$ starting from the left) will be plugged. Indeed, the energy spectrum of the superposed state, shown in Fig.~\ref{fig:lattice}(c) shows 6 states remaining in the original lowest band, but two states have been shifted lower in energy and thus out of resonance with this band. By clicking on the $\{ \psi_n\}$ checkbox, the user can verify that the two different bands correspond to states localized in the plugged and unplugged wells, respectively.

\begin{figure}[h!]
    \centering
    \includegraphics[width=6in]{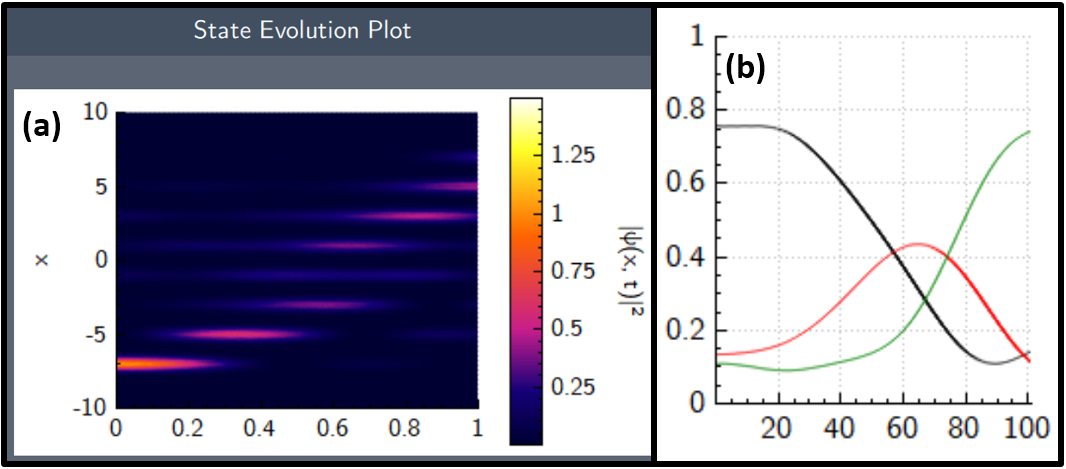}
    \caption{(a) \textit{State Evolution Plot} showing the probability density (y-axis) of the lattice wave function in the unplugged ($A=0$) potential as time evolves (x-axis). (b) \textit{Scalar Time Trace Plot} showing the total integrated population to the left of the plug (wells $1-3$, black), inside the plug (wells $4-5$, red), and on the other side of the plug (wells $6-8$, green).
    Without the plug, the wave function tunnels freely throughout the entire lattice.
    }
    \label{fig:no_plug}
    \centering
    \includegraphics[width=6in]{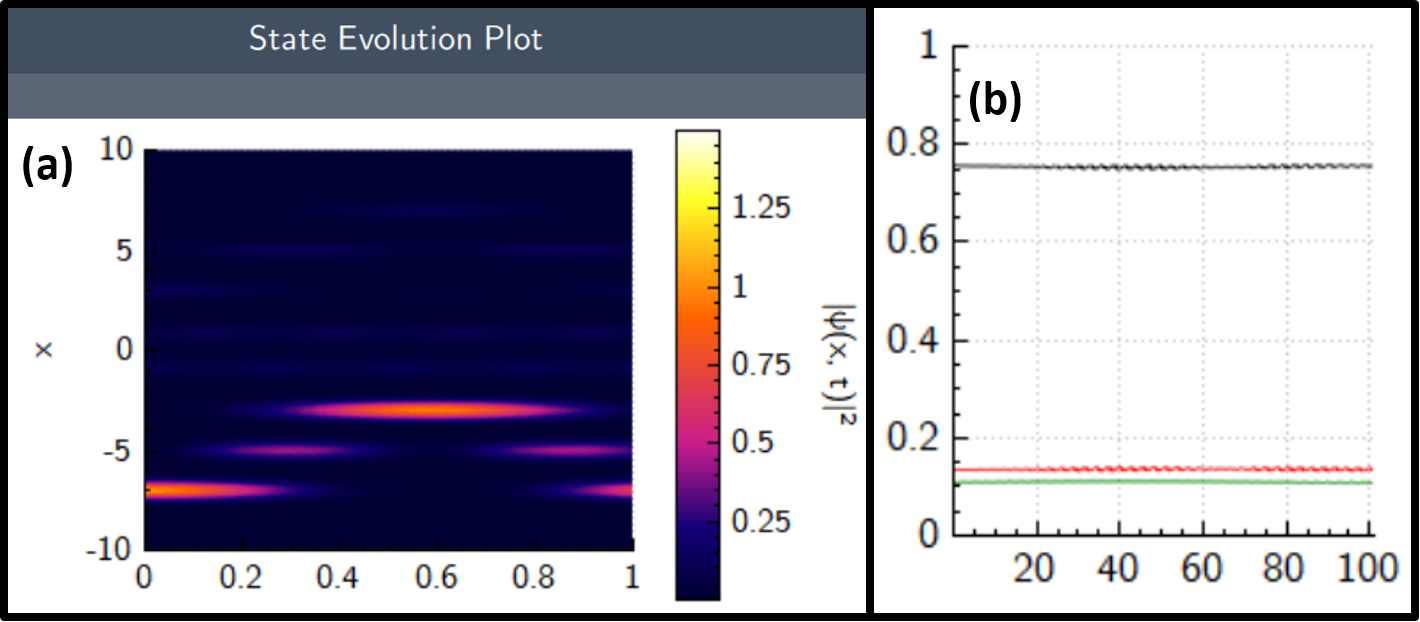}
    \caption{
    Same as Fig.~\ref{fig:no_plug}, but for the plugged potential ($A=5$). The wave function is now prohibited from tunnelling into and out of the (attractive) plug.
    }
    \label{fig:plug}
\end{figure}

The time-evolution of the state in Fig.~\ref{fig:lattice}(b) without (with) the plug are shown in Fig.~\ref{fig:no_plug} (Fig.~\ref{fig:plug}). 
Because the initial state is not completely localized to a single well, we plot the integrated population in three sections of the lattice: to the left of the plug (wells $1-3$), in the plug (wells $4-5$), and to the right of the plug (wells $6-8$). As expected, the majority of the initial state lies in wells $1-3$.

Without the plug, the wave function evolves such that it is non-negligibly present in all wells at some time, and thus the population evolves between the three sections of the lattice. This makes sense intuitively, as the localized wave function is a superposition of all $8$ states in the lowest band, and each of the $8$ eigenstates is delocalized in the lattice. With the plug, however,  the state population stays within the respective sections. Thus, the plug not only prevents population on either side from moving through to the other side, but movement into or out of the plug section is also restricted, even though it is classically attractive.

This simulation is certainly a simplified case of the actual experiment, but it allows us as experimental researchers to quickly simulate complex scenarios that move beyond simple back-of-the-envelope calculations without the use of complex numerical software. Indeed, we have experienced that what is shown here could feasibly be implemented and understood by an advanced Bachelor's student doing a project in our lab. For reference, the flowscene used in this section can be downloaded and explored at \url{www.quatomic.com/composer/exercises/}.

\section{Conclusion and Outlook}

Quantum Composer is a tool for visualizing and simulating one dimensional static and dynamic quantum systems governed by the Schr\"{o}dinger and Gross-Pitaevskii equations. The novelty of the tool is its graphical user interface and node-based design, which provides a unified, dynamic platform for teaching, learning, and researching. The drag-and-drop interface allows users to assemble nodes and create flexible flow-based simulations or adjust pre-existing simulations. The strong visual components and immediate feedback mechanisms in Composer facilitate rapid iterability and exploration.

The tool can be used for educational purposes, where instructors and students can use Composer to teach and learn quantum mechanics through visualization and exploration of potentials and dynamics of quantum states, student-driven discussions, assignments and student projects. For research purposes, we have found Composer is relevant across different project stages (e.g. in precursor studies or post-analysis) in a variety of settings, notably in quantum optimal control problems or the determination of trapping geometries relevant to our ultracold atom experiments.

Our future work will focus on studies of Composer effectiveness both from the perspective of student learning~\cite{PERC_2020} as well as its use as a research-assisting tool. We also plan to enrich the library of educational resources and expand the reach of their implementation. 

\begin{acknowledgments}

We acknowledge support from European Union's Horizon 2020 research and innovation programme under the Marie Sklodowska-Curie QuSCo grant agreement N$^\mathrm{o}$ 765267, the ERC under H2020 grant 639560 (MECTRL), and the John Templeton foundation. We would also like to acknowledge Brian Julsgaard for valuable comments in the preparation of the manuscript, and Till Heinzel and Florian Korsakissok for their work in designing and developing Quantum Composer. 

\end{acknowledgments}

\end{document}